\documentclass[prl,floatfix,twocolumn,superscriptaddress,showpacs,preprintnumbers,longbibliography,nofootinbib]{revtex4-2}
\usepackage{color}
\usepackage[utf8]{inputenc}
\usepackage[usenames,dvipsnames]{xcolor}
\usepackage{amsmath}
\usepackage{amssymb}
\usepackage{graphicx}
\graphicspath{{graphics/}}
\usepackage{epsfig}
\usepackage{dcolumn}% Align table columns on decimal point
\usepackage{bm}% bold math
\usepackage{mathrsfs}
\usepackage{multirow}
\usepackage[all]{xy}
\usepackage{pbox}
\usepackage{lipsum}
\usepackage{verbatim}
\usepackage{braket}
\usepackage{dsfont}
\usepackage{array}
\usepackage{makecell}
\usepackage{tabularx}
\usepackage{isotope}
\usepackage{xr}
\usepackage{amsthm}

\DeclareMathOperator{\Tr}{Tr}
\DeclareMathOperator{\sgn}{sgn}

\usepackage[colorlinks=true,linkcolor=blue,urlcolor=blue,citecolor=blue,pdfusetitle]{hyperref}

\begin{document}
\title{Quantum fluctuations and correlations in open quantum Dicke models}
\author{Mario Boneberg}
\affiliation{Institut f\"ur Theoretische Physik, Universit\"at Tübingen, Auf der Morgenstelle 14, 72076 T\"ubingen, Germany}
\author{Igor Lesanovsky}
\affiliation{Institut f\"ur Theoretische Physik, Universit\"at Tübingen, Auf der Morgenstelle 14, 72076 T\"ubingen, Germany}
\affiliation{School of Physics and Astronomy and Centre for the Mathematics and Theoretical Physics of Quantum Non-Equilibrium Systems, The University of Nottingham, Nottingham, NG7 2RD, United Kingdom}
\author{Federico Carollo}
\affiliation{Institut f\"ur Theoretische Physik, Universit\"at Tübingen, Auf der Morgenstelle 14, 72076 T\"ubingen, Germany}

\begin{abstract}
In the vicinity of ground-state phase transitions quantum correlations can display non-analytic behavior and critical scaling. This signature of emergent collective effects has been widely investigated within a broad range of equilibrium settings. However, under nonequilibrium conditions, as found in open quantum many-body systems, characterizing quantum correlations near phase transitions is challenging. Moreover, the impact of local and collective dissipative processes on quantum correlations is not broadly understood. This is, however, indispensable for the exploitation of quantum effects in technological applications, such as sensing and metrology. Here we consider as a paradigmatic setting the superradiant phase transition of the open quantum Dicke model and characterize quantum and classical correlations across the  phase diagram. We develop an approach to quantum fluctuations which allows us to show that local dissipation, which cannot be treated within the commonly employed Holstein-Primakoff approximation, rather unexpectedly leads to an enhancement of collective quantum correlations, and to the emergence of a nonequilibrium superradiant phase in which the bosonic and spin degrees of freedom of the Dicke model are entangled. 
\end{abstract}

\maketitle

\begin{figure}[t]
\centering
    \includegraphics[width=0.95\linewidth]{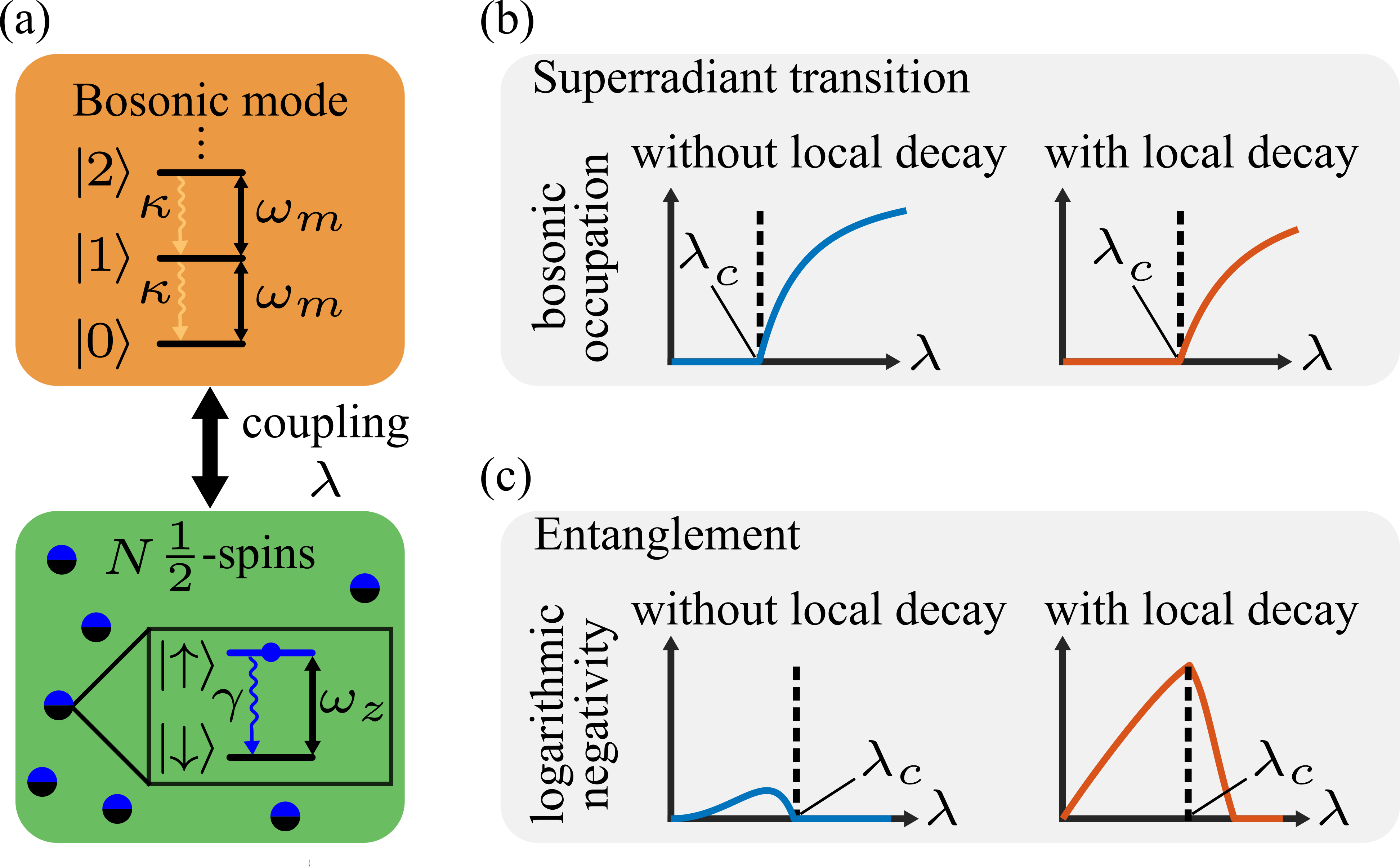}
    \caption{\textbf{Open quantum Dicke model: superradiant phase transition and entanglement.} (a) An ensemble of $N$ spin-$1/2$ systems (with energy splitting $\omega_z$ between up-state $\ket{\uparrow}$ and down-state $\ket{\downarrow}$) is coupled to a bosonic mode (the frequency $\omega_m$ determines the energy-cost of creating one field excitation $\ket{n}\rightarrow\ket{n+1}$). The presence of an environment induces boson losses (at rate $\kappa$) and local spin-decay (at rate $\gamma$). (b) At a critical coupling strength $\lambda=\lambda_c$, the system undergoes a superradiant phase transition, characterized by a macroscopic occupation ($\propto N$) of the bosonic mode, both in the presence and in the absence of local spin-decay. (c) The presence of local spin-decay leads to stronger quantum correlations and also ``stabilizes" an entangled nonequilibrium superradiant phase.}
    \label{Fig3}
\end{figure}

Spin-boson models are paradigmatic theoretical models describing, for instance, the coupling of matter with electromagnetic fields or vibrational modes. A prominent example is the so-called Dicke model \cite{dicke1954}, which provides a simple framework for the study of the interaction between a large ensemble of atoms, described by $N$ spin-1/2 (two-level) particles, and an electromagnetic cavity field, described by a bosonic mode [cf.~Fig.~\ref{Fig3}(a)]. This model has been thoroughly investigated in equilibrium  \cite{HeppLieb1,HeppLieb2,WangHioe, Hioe,DickeHigherOrderCorrections,DickeAntiresonantInteractions}, where it displays a second-order ground-state transition from a normal to a superradiant phase. While the order-parameter behavior is captured by a mean-field treatment \cite{KirtonReview}, studying quantum correlations requires the analysis of quantum fluctuations \cite{QuantumChaosTriggered,ChaosAndQPT,EntanglementAndPTSingleModeSuperradiance,EntanglementAndEntropySpinBosonQPT,MultipartiteCorrelations}. In equilibrium, this is typically done within the so-called Holstein-Primakoff approximation \cite{HP}. This exploits that the system Hamiltonian can be written in terms of collective (macroscopic) spin operators, which approximately behave as bosons when the system is close to its ground-state. 

Nowadays, also due to a debate concerning a no-go theorem on the experimental realization of the equilibrium Dicke model \cite{A2Term,nogotheorem,CoulombIntGaugeInvDM}, the focus is on the investigation of Dicke physics in an open quantum system setting [cf.~Fig.~\ref{Fig3}(a)]. Open Dicke models feature a \emph{nonequilibrium} superradiant phase transition, see Fig.~\ref{Fig3}(b), which is exactly captured by a mean-field approach \cite{SupressingRestoring,Proof}. However, in these settings, analyzing quantum fluctuations is challenging \cite{KirtonReview,KeldyshOpenSystems,KeldyshNPTQuantumOptics,buca2019}. As a consequence, very little is known about correlations in the nonequilibrium Dicke model phase transition, and even less in the presence of local dissipative processes, such as local spin-decay [cf.~Fig.~\ref{Fig3}(a)]. 

In this paper, we provide a complete characterization of quantum and classical correlations in open quantum Dicke models. We achieve this by developing an approach for treating quantum correlations, which is based on the theory of quantum fluctuation operators \cite{goderis1989a,goderis1989b,goderis1990,verbeure2010,benatti2017,benatti2018}, that can be applied also in cases where the Holstein-Primakoff approximation cannot be exploited, e.~g.~in the presence of local dissipative terms which prevent a representation of the dynamics through collective spin operators. We focus on various correlation measures, such as quantum discord and classical correlation, and show that they display non-analytic behavior at the critical coupling strength [see e.~g.~Fig.~\ref{Fig3}(c)]. Furthermore, we analyze bipartite entanglement between the spins and the bosonic mode. We find that the presence of local spin-decay --- an unavoidable process in experiments which is usually considered detrimental for quantum effects --- is unexpectedly beneficial for the build-up of quantum correlations. Our results indicate that this process leads to increased entanglement in the normal phase, and to the emergence of a nonequilibrium superradiant phase [cf.~Fig.~\ref{Fig3}(c)] where entanglement is nonvanishing.

\noindent {\bf Open quantum Dicke model.---} The Dicke model consists of an ensemble of $N$ spin-$1/2$ subsystems collectively interacting with a single bosonic mode, see Fig.~\ref{Fig3}(a). Spin operators for the $k$th particle are denoted as $\sigma^\alpha_k$, with  $\sigma^x=(\ket{\uparrow }\bra{ \downarrow }+\ket{\downarrow }\bra{\uparrow })/2$, $\sigma^z=(\ket{\uparrow} \bra{\uparrow}-\ket{\downarrow} \bra{\downarrow })/2$, and  $\sigma^y=-2i \sigma^z \sigma^x$. Here, the states $\ket{\uparrow}$, $\ket{\downarrow}$ are the single-particle spin states.  
The bosonic mode is described by creation and annihilation operators $a^\dagger$ and $a$, respectively. For later convenience, we also  introduce the spin operators $\sigma^{\pm}=\sigma^{x}\pm i \sigma^{y}$ and the bosonic quadrature operators $q=i(a-a^{\dagger})/\sqrt{2}$ and $p=(a+a^{\dagger})/\sqrt{2}$. 

The system is governed by a Markovian open quantum dynamics, under which the time-evolution of an operator $O$ follows the Lindblad equation $\dot{O}(t)=\mathcal{L}_N[O(t)]$ \cite{TheoryOQS,GeneratorsQuantumDynamicalSemigroups,CompletelyPositive} with generator
\begin{align}\label{Lindbladian}
\mathcal{L}_N[O]:&=i[H_N^D,O] + \kappa \left( a^{\dagger} O a - \frac{1}{2} \{ a^{\dagger} a ,O \} \right) \nonumber\\
&+\gamma \sum_{k=1}^N\left(\sigma_k^+O\sigma_k^- -\frac{1}{2}\{\sigma_k^+ \sigma_k^-,O\}\right)\, .
\end{align}
The first term on the right-hand side of Eq.~\eqref{Lindbladian} gives the coherent contribution to the dynamics implemented by the Dicke Hamiltonian (setting $\hbar=1$)
\begin{align}\label{H_N^D}
H_N^D=\omega_m a^{\dagger} a + \omega_z S^z + \frac{2 \lambda}{\sqrt{N}} (a+a^{\dagger}) S^x\, .
\end{align}
Here, $\omega_m>0$ is the bosonic mode frequency, $\omega_z>0$ the energy splitting between spin states and $\lambda>0$ the coupling parameter  [cf.~Fig.~\ref{Fig3}(a)]. The Dicke Hamiltonian \eqref{H_N^D} is written in terms of the collective operators $S^{\alpha}= \sum^N_{k=1} \sigma^{\alpha}_k$, obeying $[S^{\alpha},S^{\beta}]=i\sum_\gamma \epsilon^{\alpha \beta \gamma}S^{\gamma}$, where $\epsilon^{\alpha \beta \gamma}$ is the Levi-Civita symbol.
The factor $1/\sqrt{N}$, which rescales the collective spin-boson coupling in $H_N^D$, is necessary for a well-defined thermodynamic limit \cite{KirtonReview}. The last two terms in Eq.~\eqref{Lindbladian} account for irreversible dynamical effects. These are decay of bosonic excitations at rate $\kappa$ as well as local (individual) spin-decay, $\ket{\uparrow}\to\ket{\downarrow}$, at rate $\gamma$. As becomes clear from Eq.~\eqref{Lindbladian}, the latter process is not described through collective, but rather local, spin (jump) operators $\sigma^-_k$. 

\noindent {\bf Superradiant transition and mean-field results.---} The open quantum Dicke model undergoes a phase transition --- as a function of the coupling strength $\lambda$ --- from a normal stationary phase, with subextensive (in $N$) bosonic occupation, to a superradiant one where the bosonic mode becomes macroscopically occupied \cite{SupressingRestoring,KirtonReview,Proof} [see sketch in Fig.~\ref{Fig3}(b)]. An order parameter for this transition is the stationary expectation of the renormalized number operator $ a^\dagger a /N $ in the thermodynamic limit of large number of spins,  $N\to\infty$. %We now discuss how this transition can be characterized.  

\begin{figure*}[t]
\centering
    \includegraphics[width=\textwidth]{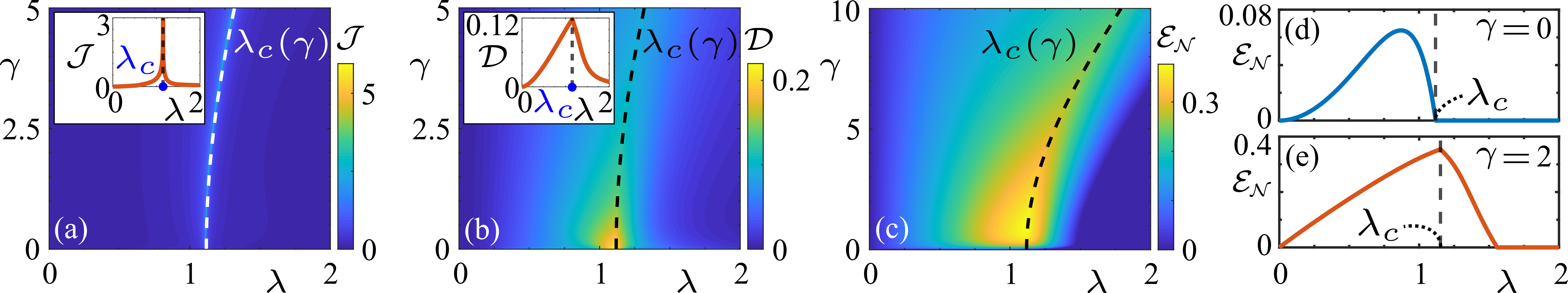}
    \caption{\textbf{Quantum and classical correlations between the spin ensemble and the bosonic mode.} (a-b) Classical correlation $\mathcal{J}$ and quantum discord $\mathcal{D}$ as functions of $\gamma$ and $\lambda$. The critical line $\lambda_c(\gamma)$ (dashed line) separates the normal phase from the superradiant one. Both quantities display a non-analytic behavior at the critical line, with the classical correlations diverging. The insets visualize the $\lambda$-dependence of the corresponding quantities for $\gamma=2$. (c) Logarithmic negativity $\mathcal{E}_{\mathcal{N}}$ as a function of $\gamma$ and $\lambda$.  (d-e) Logarithmic negativity $\mathcal{E}_{\mathcal{N}}$ as a function of $\lambda$ for  $\gamma=0$ (blue) and $\gamma=2$ (red). For all plots, we fixed $\omega_m=1$ and $\omega_z=4$. All parameters are given in units of $\kappa$.}
    \label{Fig1}
\end{figure*}

The form of the order parameter suggests the definition of the so-called \textit{mean-field operators}
\begin{align} \label{xp}
m_N^q=\frac{q}{\sqrt{N}},\quad  m_N^p=\frac{p}{\sqrt{N}}\quad  \textrm{and} \quad m_N^{\alpha}=\frac{S^{\alpha}}{N}\, , 
\end{align}
where the last term must be considered for $\alpha=x,y,z$. The first two operators are relevant as they provide the order parameter through $ (m_N^q)^2+(m_N^p)^2-1/N=2a^\dagger a /N$. The mean-field operators of the spin ensemble [last terms in Eqs.~\eqref{xp}] are also important for studying the model. Indeed, by computing the action of the generator $\mathcal{L}_N$ in  Eq.~\eqref{Lindbladian} on the mean-field operators in Eqs.~\eqref{xp}, one finds that these operators are all dynamically coupled \cite{Proof}. In the thermodynamic limit, the time-evolved operators $m_N^\alpha(t)$ (with $\alpha=x,y,z,q,p$) behave as multiples of the identity proportional to their expectation, i.~e.~$m^\alpha_N(t)\to  m^\alpha (t)=\lim_{N\to\infty}\langle m_N^\alpha(t)\rangle$ \cite{SupressingRestoring,Proof}. Furthermore, they obey the differential equations (we drop the explicit time-dependence)
\begin{align} 
  \dot{m}^{x} &= - \omega_z  m^{y}   -\frac{\gamma}{2}   m^x  \, ,\nonumber \\
  \dot{m}^{y}   &=  \omega_z  m^{x}   - 2^{3/2} \lambda     m^p    m^{z}  -\frac{\gamma}{2}  m^y\, , \nonumber\\
 \dot{m}^{z}  &=  2^{3/2} \lambda m^p   m^{y}  -\frac{\gamma}{2} (1+2  m^z  ) \, ,\label{mnfld}\\
\dot{m}^q  &= \omega_m m^p  + 2^{3/2}  \lambda  m^x  -  \frac{\kappa}{2}  m^q  \, ,\nonumber\\
 \dot{m}^p &= - \omega_m  m^q  -  \frac{\kappa}{2} m^p \, . \nonumber
\end{align}
These equations feature two different stationary regimes, separated by a critical value of the coupling strength
\begin{equation}
    \lambda_c=\sqrt{\frac{\left[\omega_m^2+\left(\frac{\kappa}{2}\right)^2\right]\left[\omega_z^2+\left(\frac{\gamma}{2}\right)^2\right]}{4 \omega_z \omega_m}}\, .
\end{equation}
For $\lambda<\lambda_c$, there exists a unique stable stationary solution to the system in Eqs.~\eqref{mnfld}, with the only nonzero value given by $ m^z=-1/2$. This is the normal phase. For $\lambda>\lambda_c$, this becomes unstable but two other stable solutions emerge which spontaneously break the symmetry $a\to -a$, $\sigma^-\to -\sigma^-$ of the generator $\mathcal{L}_N$ \cite{KirtonReview,Dickemodel,SM}.\vphantom{\cite{Hurwitz,NonlinearSystems,Gantmacher,MatrixAnalysis,LyapEqu,SimonMukundaDutta,simon1987}} These feature finite stationary values of $ m^{x/p}$, which imply a macroscopic occupation of the bosonic mode in this superradiant phase. (Details are provided in the Supplemental Material \cite{SM}.) 

\noindent {\bf Quantum fluctuations.---} The observables $m^\alpha_N$ provide, in the thermodynamic limit, $N\to\infty$, a classical (mean-field) description of the Dicke model \cite{lanford1969,benatti2018,Proof} which carries no information about correlations. In order to go beyond this, we introduce a new set of observables, so-called quantum \textit{fluctuation operators} \cite{goderis1989a,goderis1989b,goderis1990,verbeure2010,benatti2017,benatti2018}. These will allow us to explore fluctuations and quantum correlations in the two stationary phases.

The fluctuation operators read
\begin{align}\label{flucalpha}
F_N^{\alpha}=\sqrt{N}\left(m_N^\alpha-\langle m_N^\alpha\rangle \right)\, .
\end{align}
For $\alpha=x,y,z$, these are the usual spin fluctuation operators \cite{benatti2017,benatti2018}, and we have defined the bosonic ones  ($\alpha=q,p$) in full analogy.
Roughly speaking, the operators in Eqs.~\eqref{flucalpha} account for deviations of the operators $m_N^\alpha$ from the mean-field behavior. Remarkably, despite being collective, these retain a quantum character in the thermodynamic limit, in which the limiting operators $F^\alpha=\lim_{N\to\infty}F^\alpha_N$ behave as bosons (for a rigorous discussion see e.~g.~Ref.~\cite{verbeure2010}). This is straightforward to check for $F^{q/p}$, since $F_N^q=q-\langle q\rangle$ and $F_N^p=p-\langle p\rangle$. However, also collective spin fluctuations give rise to an emergent bosonic mode. This can be seen as follows. Looking at the commutator of fluctuation operators, one finds that  $[F^\alpha,F^\beta]=i\sum_\gamma  \epsilon^{\alpha \beta \gamma} m^{\gamma}$ ($\alpha,\beta=x,y,z$), which is a multiple of the identity. Now, we rotate the reference frame for the spin ensemble aligning the $z$-direction with the direction identified by mean-field variables, which is $\hat{n}=  \vec{m}^s  /| \vec{m}^s |$ with $\vec{m}^s=(m^x,m^y,m^z)^T$. In this rotated frame we have $\tilde{m}^x=\tilde{m}^y=0$ and $\tilde{m}^z>0$, so that the only nonzero commutator is $[\tilde{F}^x,\tilde{F}^y]=i\tilde{m}^z$. A \textit{canonical} bosonic mode is finally obtained by defining $Q=\tilde{F}^x/\sqrt{\tilde{m}^z}$, $P=\tilde{F}^y/\sqrt{\tilde{m}^z}$, which fulfill  $[Q,P]=i$. 

In what follows, we work with the set of fluctuations $r=(Q,P,\tilde{F}^z,F^q,F^p)^T$. The first two elements represent an emergent bosonic mode describing collective properties of the spin ensemble; the last two are the fluctuations of the original bosonic mode, while $\tilde{F}^z$ is a fluctuation operator which commutes with the others  \cite{verbeure2010,benatti2018}. 

To analyze correlations in the Dicke model through fluctuation operators, we introduce the covariance matrix $\tilde{\Sigma}^{\alpha \beta}=\langle \{r^{\alpha} ,r^{\beta} \} \rangle/2$. For Gaussian states, this matrix contains the full information about fluctuations and can even be used to quantify collective correlations \cite{benatti2014,benatti2016c}. Before going to that, however, we briefly discuss the time-evolution of $\tilde{\Sigma}$ under the dynamics implemented by the generator in Eq.~\eqref{Lindbladian}. For each parameter regime, we consider the dynamics of fluctuations emerging, in the thermodynamic limit, from an initial state which is stationary with respect to the mean-field observables and possesses Gaussian fluctuations. In this setting, the covariance matrix obeys the differential equation \cite{benatti2018,Feedback}
\begin{align} \label{CM}
\dot{\tilde{\Sigma}}(t)=\tilde{\Sigma}(t) \tilde{G}^T+\tilde{G}\tilde{\Sigma}(t) +\tilde{W},
\end{align}
where the matrices $\tilde{G}$ and $\tilde{D}$, whose explicit form is given in \cite{SM}, depend on the parameters of the model and on the stable stationary mean-field variables of Eqs. (\ref{mnfld}). The time-evolution in Eq.~\eqref{CM} has the structure of a bosonic Gaussian open quantum dynamics \cite{heinosaari2010}, suggesting that the Gaussianity of fluctuations is preserved at all times. Moreover, as long as $\lambda\neq\lambda_c$, Eq.~\eqref{CM} has a unique stationary solution $\tilde{\Sigma}_\infty$ \cite{SM}. 

Since we are mainly interested in quantum correlations, we discard the information associated with the trivial fluctuation $\tilde{F}^z$. This can be done by extracting from the stationary covariance matrix $\tilde{\Sigma}_\infty$ the $4\times4$ minor obtained by neglecting its  third row and its third column. The resulting covariance matrix 
\begin{align*}
\tilde{\Sigma}^{\mathrm{t-m}}_\infty
= \begin{pmatrix}
\Gamma_s & \Gamma_c \\
\Gamma_c^T & \Gamma_b 
\end{pmatrix}\, ,
\end{align*}
contains the full information about the two bosonic modes $Q,P$ and $F^q,F^p$. In particular,  $\Gamma_s$ is the $2\times2$ matrix containing the second moments of the operators $Q,P$, $\Gamma_b$ contains those of $F^q,F^p$, and $\Gamma_c$ contains correlations between $Q,P$ and $F^q,F^p$.  

\noindent {\bf Quantum and classical correlations.---} In order to explore the correlation structure in the open quantum Dicke model, we focus on measures which can distinguish between correlations of different nature, e.~g.~quantum or classical, and that are fully determined by the covariance matrix $\tilde{\Sigma}^{\mathrm{t-m}}_\infty$ \cite{SM}. Since the spin fluctuation operators involve all the spin degrees of freedom, the correlations that we discuss here are of collective type, i.~e.~reflecting how the spin ensemble as a whole is collectively correlated with the bosonic mode. 

Firstly, we consider the \textit{classical correlation} $\mathcal{J}$ \cite{adesso2010, giorda2010, isar2014, henderson2001, ollivier2001, SM} between the spin ensemble and the bosonic mode. This quantity encodes the maximum information that can be extracted on one subsystem, by making generalized (Gaussian) measurements on the other one. In this sense, the classical correlation is ``asymmetric" since it can be defined in two ways, i.~e.~either considering that measurements are performed on the spin ensemble or on the bosonic mode. Secondly, we study the so-called \textit{quantum discord} $\mathcal{D}$ \cite{adesso2010, giorda2010, isar2014, henderson2001, ollivier2001, SM}, which is defined as the difference between the total correlation --- quantified by the quantum mutual information --- and the classical correlation $\mathcal{J}$. This quantity measures the genuine quantum contribution to the total correlation between the two subsystems. According to its definition through the classical correlation, also the quantum discord is asymmetric under exchange of the role of the spin ensemble and of the bosonic mode. In the following, we consider both quantum discord and classical correlation assuming that the measurements are performed on the bosonic mode (see results in \cite{SM} for the other case). 

In Fig.~\ref{Fig1}(a-b), we show the stationary behavior of classical correlation and quantum discord, as a function of the coupling strength $\lambda$ and of the local spin-decay rate $\gamma$. As shown, the classical correlation diverges at the nonequilibrium phase transition line, witnessing strong spin-boson correlations. Concerning the presence of quantum correlations, we observe that quantum discord $\mathcal{D}$ is different from zero almost everywhere in the phase diagram. It is maximal along the critical line, where it shows a non-analytic behavior even though it remains bounded. 

We now consider the emergence of collective entanglement between the spins and the bosonic mode. This can be quantified from the covariance matrix $\tilde{\Sigma}^{\mathrm{t-m}}_\infty$, through the \textit{logarithmic negativity} $\mathcal{E}_{\mathcal{N}}$ --- a proper entanglement measure --- defined as \cite{vidal2002,plenio2005,AdessoIlluminatiTopicalReview, SM}
\begin{align*}
\mathcal{E}_{\mathcal{N}}=\max (0,-\log ( \tilde{\nu}_- ))\, .
\end{align*}
Here, $\tilde{\nu}_-$ is the smallest (symplectic) eigenvalue \cite{AdessoIlluminatiTopicalReview} of the partially transposed covariance matrix obtained from $2\tilde{\Sigma}^{\mathrm{t-m}}_\infty$ by exchanging $F^p\to -F^p$ \cite{SimonMukundaDutta,simon1987,Williamson,SimonPPT}. As we show in Fig.~\ref{Fig1}(c), the open quantum Dicke model displays collective spin-boson entanglement in a large parameter regime. We are particularly interested in understanding the impact of local spin-decay on entanglement. For small, yet nonvanishing, values of $\gamma$ we identify a pronounced peak near the critical line $\lambda_c(\gamma)$. This suggests that a small rate of local spin-decay leads to larger entanglement. However, when $\gamma$ vanishes, entanglement is dramatically reduced. This becomes evident when comparing the behavior of entanglement, as a function of $\lambda$, for $\gamma=0$ and $\gamma\neq0$. An example is shown in Fig.~\ref{Fig1}(d-e). In the absence of local spin-decay, entanglement vanishes at the critical point and is always zero in the superradiant phase [cf.~Fig.~\ref{Fig1}(d)]. However, when local spin-decay is present, entanglement assumes larger values across the whole phase diagram and can also persist in the superradiant phase [cf.~Fig.~\ref{Fig1}(e)]. Furthermore, for $\gamma\neq0$, the logarithmic negativity shows a non-analytic behavior at the critical point and undergoes a ``sudden death"  well inside the superradiant phase, as shown in Fig.~\ref{Fig1}(e).  These results show that local spin-decay has, rather surprisingly, an overall beneficial effect on quantum correlations, and on quantum entanglement in particular. Comparing Fig.~\ref{Fig1}(b) and Fig.~\ref{Fig1}(c), we also see that there exist parameter regions where the quantum discord assumes a finite value but the logarithmic negativity is zero. In this region, the quantum state of fluctuations is separable but nevertheless non-trivially quantum correlated. 

\begin{figure}[t]
\centering
    \includegraphics[width=8.6cm]{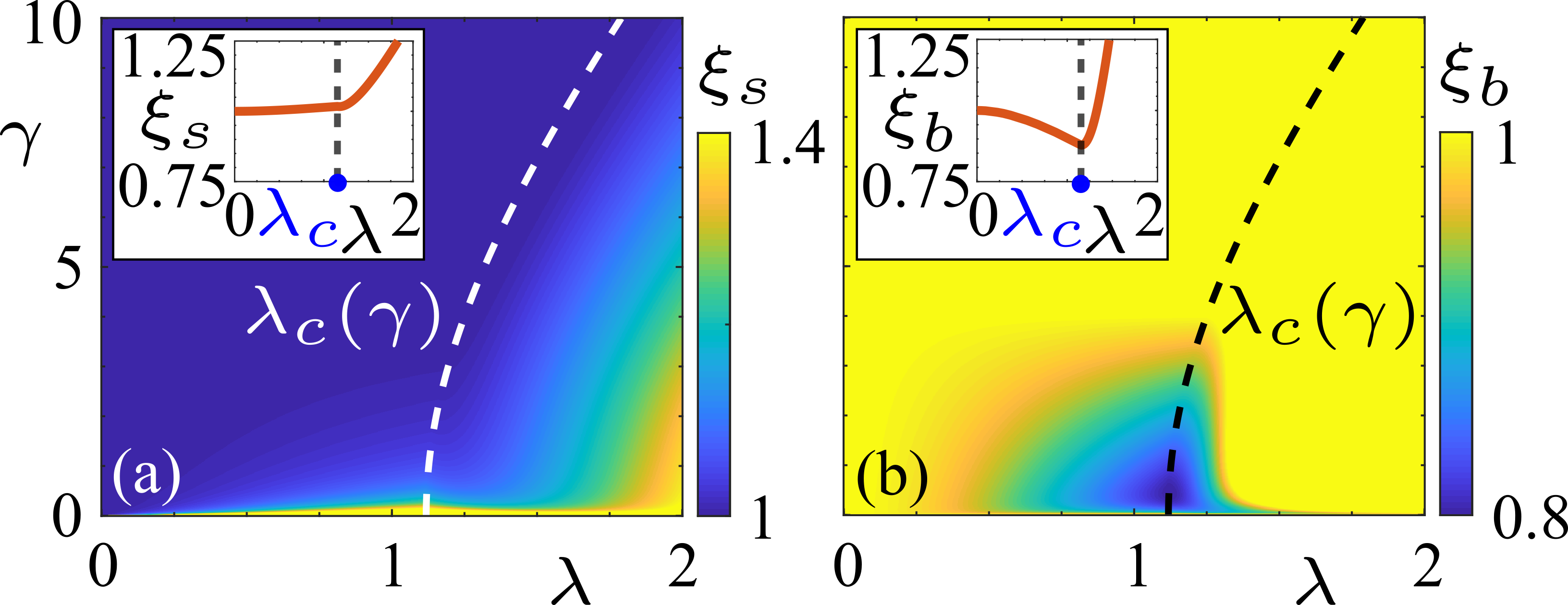}
    \caption{\textbf{Squeezing.} (a) Spin squeezing parameter $\xi_s$ and (b) boson squeezing parameter $\xi_b$ as functions of $\gamma$ and $\lambda$. Both insets show a cut through the density plot at $\gamma=2$. We have chosen $\omega_m=1$ and  $\omega_z=4$. All parameters are in units of $\kappa$.}
    \label{Fig2}
\end{figure}

Finally, we analyze quantum correlations within each subsystem separately. These are measured by the squeezing parameter $\xi=2\min(\Theta_1,\Theta_2)$ \cite{SSS,AtomicCoherentStates,SSGross,Feedback} where $\Theta_1$, $\Theta_2$ denote the eigenvalues of $\Gamma_s$ for spin squeezing and of $\Gamma_b$ for boson squeezing. The parameter $\xi$ quantifies the minimum variance among all possible quadrature operators. A state is called squeezed if $\xi<1$, i.~e.~if the variance in one of the quadratures is smaller than the smallest possible simultaneous uncertainty of two canonically conjugated quadrature operators (also referred to as shot-noise limit \cite{SSGross}).
Fig.~\ref{Fig2}(a) shows that there is no spin-squeezing in the stationary state of the model since $\xi_s$ is always larger than or equal to $1$. In contrast, the system can feature squeezing in the bosonic mode below a threshold, i.~e.~$\gamma \lesssim 4$. As shown in the inset of Fig. \ref{Fig2}(b) the bosonic squeezing parameter $\xi_b$ takes its minimum values near $\lambda=\lambda_c(\gamma)$.\\

\noindent {\bf Discussion.---} We explored the stationary structure of correlations in an open quantum Dicke model. We found that, in the absence of local spin-decay ($\gamma=0$), the superradiant phase does not feature spin-boson entanglement. Even though this may appear somehow counter-intuitive since superradiant phases arise in the strong coupling regime \cite{ControlMultiatomEntanglement}, disentangled nonequilibrium superradiant phases have also been observed in other settings \cite{CertifyingSeparability}. However, as we have shown, the presence of local spin-decay ($\gamma\neq0$) appears to be beneficial for the build-up of quantum correlations and can even be used to ``stabilize" entanglement in superradiant stationary regimes. Furthermore, through other measures of correlations, we have shown that, even when there is no spin-boson entanglement, there are residual quantum correlations in the system which evidence non-classical properties across the whole phase diagram of the open quantum Dicke model. Also these correlations, and not only entanglement, could be exploited to achieve quantum-enhanced sensitivity in metrological applications \cite{braun2018}.

\begin{acknowledgements}
\noindent \textbf{Acknowledgments.} We acknowledge support from the ``Wissenschaftler R\"{u}ckkehrprogramm GSO/CZS" of the Carl-Zeiss-Stiftung and the German Scholars Organization e.V., the Deutsche Forschungsgemeinschaft through Grant No. 435696605 and 449905436, as well as from the Baden-W\"urttemberg Stiftung through project BWST\_ISF2019-023.
\end{acknowledgements}

\bibliography{Reference}

\setcounter{equation}{0}
\setcounter{figure}{0}
\setcounter{table}{0}
\makeatletter
\renewcommand{\theequation}{S\arabic{equation}}
\renewcommand{\thefigure}{S\arabic{figure}}

\makeatletter
\renewcommand{\theequation}{S\arabic{equation}}
\renewcommand{\thefigure}{S\arabic{figure}}

\onecolumngrid
\newpage

\begin{center}
{\Large SUPPLEMENTAL MATERIAL}
\end{center}
\begin{center}
\vspace{0.8cm}
{\Large Quantum fluctuations and correlations in open quantum Dicke models}
\end{center}
\begin{center}
Mario Boneberg,$^{1}$ Igor Lesanovsky,$^{1,2}$ and Federico Carollo$^1$
\end{center}
\begin{center}
$^1${\em Institut f\"ur Theoretische Physik, Universit\"at T\"ubingen,}\\
{\em Auf der Morgenstelle 14, 72076 T\"ubingen, Germany}\\
$^2$ {\em School of Physics and Astronomy and Centre for the Mathematics}\\
{\em and Theoretical Physics of Quantum Non-Equilibrium Systems,}\\
{\em The University of Nottingham, Nottingham, NG7 2RD, United Kingdom}
\end{center}

\section{Mean-field analysis}
In this section, we show how the mean-field equations, Eqs.~\eqref{mnfld}, can be derived from the Lindblad generator $\mathcal{L}_N$. We then obtain the stationary solutions for such equations and analyze their stability. Summation over repeated indices is implied here and in the following.

\subsection{Derivation of the mean-field equations}

Starting point for the derivation of Eqs.~\eqref{mnfld} is the superoperator in Eq.~\eqref{Lindbladian}, which can be rewritten as \begin{align*}
\mathcal{L}_N[X]=i[H_N^D,X] + \underbrace{\frac{\kappa}{2} ([a^{\dagger},X]a +a^{\dagger} [X,a])}_{=:\mathcal{D}_N^{\kappa}[X]}+ \underbrace{ \frac{\gamma}{2} \sum_{k=1}^N([\sigma_k^+,X]\sigma_k^- +\sigma_k^+[X,\sigma_k^-])}_{=:\mathcal{D}_N^{\gamma}[X]},
\end{align*}
where we have used that for an arbitrary operator $A$ 
\begin{align*}
A^{\dagger}XA -\frac{1}{2} \{ A^{\dagger} A,X \} = \frac{1}{2} ([A^{\dagger},X]A +A^{\dagger} [X,A]).
\end{align*}
We separate the problem of calculating the action of $\mathcal{L}_N$ on the mean-field operators, by looking separately at the actions of the contributions $i[H_N^D,X]$, $\mathcal{D}_N^{\kappa}[X]$ and $\mathcal{D}_N^{\gamma}[X]$. For this, we note that 
\begin{align*}
    \mathcal{D}_N^{\kappa}[X]= \frac{N \kappa}{4} ([m_N^p+im_N^q,X](m_N^p-im_N^q) +(m_N^p+im_N^q) [X,m_N^p-im_N^q])
\end{align*}
and 
\begin{align} \label{H_N^Dcommutator}
    i[H_N^D,X]=i[\frac{N\omega_m}{2} ({m_N^q}^2+{m_N^p}^2) + \omega_z S^z + 2^{3/2} \lambda m_N^p S^x,X].
\end{align}
From the latter we get 
\begin{align*}
    i[H_N^D,m_N^{\alpha}] = - \omega_z \epsilon^{z \alpha \gamma} m_N^{\gamma} - 2^{3/2} \lambda m_N^p \epsilon^{x \alpha \gamma} m_N^{\gamma},
\end{align*}
for $\alpha=x,y,z$, as well as 
\begin{align*}
        i[H_N^D,m_N^q]=\omega_m m_N^p  + 2^{3/2} \lambda m_N^x \quad \textrm{and} \quad
        i[H_N^D,m_N^p]= -\omega_m m_N^q.
\end{align*}
For the \textit{dissipator} $\mathcal{D}_N^{\kappa}$, we immediately see that $\mathcal{D}_N^{\kappa}[m_N^{\alpha}]=0$, for $\alpha=x,y,z$, and that  
\begin{align*}
    \mathcal{D}_N^{\kappa}[m_N^q]=-\frac{\kappa}{2}m_N^q, \quad
    \mathcal{D}_N^{\kappa}[m_N^p]=-\frac{ \kappa}{2}m_N^p.
\end{align*}
For the dissipator $\mathcal{D}_N^{\gamma}$, in contrast, we have that  $\mathcal{D}_N^{\gamma}[m_N^q]=0$ and $\mathcal{D}_N^{\gamma}[m_N^p]=0$. Using the identities $[\sigma_k^{\alpha},\sigma_i^{\beta}]=i\delta_{ki}\epsilon^{\alpha \beta \gamma} \sigma_i^{\gamma}$, $\sigma_k^{x}=\frac{\sigma_k^+ +\sigma_k^-}{2}$, $\sigma_k^y=\frac{\sigma_k^+-\sigma_k^-}{2i}$ and $\sigma_k^{\rho}\sigma_k^{\nu}=\delta^{\rho \nu}\frac{\mathds{1}_k}{4}+i \epsilon^{\rho \nu \mu} \frac{\sigma_k^{\mu}}{2}$ it also follows that
\begin{align*}
\mathcal{D}_N^{\gamma}[m_N^{\alpha}]&=\frac{\gamma}{2N} \sum_{i=1}^N(\delta^{x\alpha}  (\{ \sigma_i^x,\sigma_i^z\} +i [\sigma_i^y,\sigma_i^z])+i\delta^{y\alpha} ([ \sigma_i^z,\sigma_i^x] -i \{\sigma_i^y,\sigma_i^z \})-2 \delta^{z\alpha}  (\frac{\mathds{1}_i}{2} -i[\sigma_i^x,\sigma_i^y]))\\
&= \frac{\gamma}{2} (\delta^{x\alpha}  (-m_N^x)+\delta^{y\alpha} (-m_N^y)+ \delta^{z\alpha}  (-\mathds{1} -2 m_N^z)).
\end{align*}

Now, considering that $\dot{m}_N^\alpha=\mathcal{L}_N[m_N^\alpha]$, taking the expectation value and using that in the large $N$ limit $m_N^\alpha\to m^\alpha$ which are proportional to the identity, one obtains the mean-field equations reported in  Eqs.~\eqref{mnfld}.

\subsection{Stationary state solution of the mean-field equations}
Setting the time derivatives of the equations in \eqref{mnfld} to zero, we get the system of equations 
\begin{align}
 m^y  &= -\frac{\gamma}{2 \omega_z}  m^x  \label{eqn:statmnfld1} \\
0 &=  \omega_z  m^x  - 2^{3/2} \lambda   m^p   m^z  -\frac{\gamma}{2}  m^y  \label{eqn:statmnfld2}\\
0 &=  2^{3/2} \lambda  m^p   m^y  -\frac{\gamma}{2} (1+2  m^z ) \label{eqn:statmnfld3}\\
0 &= \omega_m  m^p  + 2^{3/2}  \lambda  m^x   -  \frac{\kappa}{2}  m^q  \label{eqn:statmnfld4}\\
 m^q  &=  -  \frac{\kappa}{2 \omega_m}  m^p  , \label{eqn:statmnfld5}
\end{align}
which determines the stationary solution. Through several substitutions we can find  
\begin{align*}
 m^z =-\frac{(\omega_m^2+(\frac{\kappa}{2})^2)(\omega_z^2+(\frac{\gamma}{2})^2)}{2^3 \lambda^2 \omega_z \omega_m}=-\frac{1}{2} \frac{\lambda_c^2}{\lambda^2} \quad \textrm{with} \quad
\lambda_c^2=\frac{(\omega_m^2+(\frac{\kappa}{2})^2)(\omega_z^2+(\frac{\gamma}{2})^2)}{4 \omega_z \omega_m}.
\end{align*}
Including also Eq. (\ref{eqn:statmnfld3}), we further extract 
\begin{align*}
 m^x &=\pm \sqrt{\frac{ \frac{\omega_z}{\omega_m}(\omega_m^2 +(\frac{\kappa}{2})^2 )}{2^{3} \lambda^2 }} \sqrt{1-\frac{\lambda_c^2}{\lambda^2} }.
\end{align*}
Thus with Eqs. (\ref{eqn:statmnfld1}), (\ref{eqn:statmnfld3}) and (\ref{eqn:statmnfld5}) we see that for $\gamma > 0$
\begin{align}
 m^x  &=\pm\frac{\omega_z}{\sqrt{2(\omega_z^2+(\frac{\gamma}{2})^2)}}\frac{\lambda_c}{\lambda}\sqrt{1-\frac{\lambda_c^2}{\lambda^2} }\nonumber\\
 m^y &=\mp \frac{\gamma}{2^{3/2} \sqrt{\omega_z^2+(\frac{\gamma}{2})^2}} \frac{\lambda_c}{\lambda} \sqrt{1-\frac{\lambda_c^2}{\lambda^2} }\nonumber\\
 m^z &=-\frac{1}{2}\frac{\lambda_c^2}{\lambda^2} \label{gammafinitesup}\\
 m^q &=\pm \frac{\kappa \omega_z\lambda_c}{(\omega_m^2+(\frac{\kappa}{2})^2)\sqrt{\omega_z^2+(\frac{\gamma}{2})^2}} \sqrt{1-\frac{\lambda_c^2}{\lambda^2} }\nonumber\\
 m^p &=\mp \frac{2 \omega_z\lambda_c}{(\omega_m+(\frac{\kappa}{2})^2\frac{1}{ \omega_m})\sqrt{\omega_z^2+(\frac{\gamma}{2})^2}}\sqrt{1-\frac{\lambda_c^2}{\lambda^2} }\nonumber
\end{align}
are stationary solutions for $\lambda\geq \lambda_c$. For $\lambda\geq 0$ there exists another stationary solution, namely 
\begin{align}
 m^x  &=0\nonumber\\
 m^y &=0\nonumber\\
 m^z &=-\frac{1}{2}\label{sub}\\
 m^q &=0\nonumber\\
 m^p &=0.\nonumber
\end{align} 

Considering the case $\gamma=0$, the latter solution is still a valid solution whenever $\lambda\geq 0$ (together with the solution having $ m^z =1/2$) but the non-trivial solutions are now given by 
\begin{align}
  m^{x}  &= \mp \frac{   \sqrt{ 1 - \frac{\lambda_c^4}{\lambda^4}} }{2}\nonumber\\
   m^{y}  &= 0 \nonumber\\
    m^{z}  &= \pm \frac{\lambda_c^2}{2 \lambda^2} \label{gammazerosup}\\
 m^q  &= \mp \frac{\kappa}{\sqrt{2}} \frac{ \lambda}{(\omega_m^2 + (\frac{\kappa}{2})^2 )} \sqrt{ 1 - \frac{\lambda_c^4}{\lambda^4}} \nonumber\\
  m^p  &= \pm \frac{ \lambda}{(\omega_m + (\frac{\kappa}{2})^2 \frac{1}{\omega_m} )/\sqrt{2}} \sqrt{ 1 - \frac{\lambda_c^4}{\lambda^4}}.\nonumber
\end{align}
for $\lambda \geq \lambda_c$. Since the sign choice for $ m^{z} $ is independent of the others, these are four solutions and the critical coupling $\lambda_c$ is the same as above, evaluated at $\gamma=0$.

\subsection{Stability analysis of the stationary mean-field solutions}
For the solutions of the last section we perform a stability analysis using Lyapunov's indirect method \cite{NonlinearSystems}. For $\gamma=0$ the constraint $ {m^z}^2=\frac{1}{4}-  {m^y}^2 -{m^x}^2 $, occuring due to the conservation of $S^2$, reduces Eqs. (\ref{mnfld}) to a system of four coupled first-order non-linear differential equations 
\begin{align*}
\dot{ m^{x} } &= -\omega_z m^y \\
\dot{ m^{y} } &= \omega_z m^x \mp 2^{3/2} \lambda m^p \sqrt{\frac{1}{4} -{m^y}^2 -{m^x}^2}\\
\dot{ m^q} &= \omega_m m^p + 2^{3/2}  \lambda m^x  -  \frac{\kappa}{2} m^q\\
\dot{ m^p} &= - \omega_m  m^q -  \frac{\kappa}{2} m^p
\end{align*}
which completely determines the dynamics. It can be written in the form $\dot{\vec{u}}=f(\vec{u})$ with $\vec{u}=(m^x,m^y,m^q,m^p)^T$ and the Jacobian of $f$ is 
\begin{align*}
J(\vec{u})= \begin{pmatrix}
 0 & -\omega_z  & 0 & 0 \\
 \omega_z \pm 2^{3/2} \lambda m^p \frac{m^x}{\sqrt{\frac{1}{4} -{m^y}^2 -{m^x}^2}} & \pm 2^{3/2} \lambda m^p \frac{m^y}{\sqrt{\frac{1}{4} -{m^y}^2 -{m^x}^2}} & 0 & \mp 2^{3/2} \lambda \sqrt{\frac{1}{4} -{m^y}^2 -{m^x}^2}\\
  2^{3/2} \lambda & 0 & -\frac{\kappa}{2} & \omega_m \\
 0 & 0 & -\omega_m & -\frac{\kappa}{2}
\end{pmatrix}.
\end{align*}
The characteristic polynomial is calculated as
\begin{align*}
    \det( \mathrm{J}(\vec{u})- \mathds{1} \chi)=&\chi^4 +(\kappa + \frac{2^{3/2} m^y m^p \lambda}{\sqrt{\frac{1}{4} -{m^x}^2 -{m^y}^2}}) \chi^3 +(\frac{\kappa^2}{4}+\frac{2^{3/2} m^y m^p \kappa \lambda}{\sqrt{\frac{1}{4} -{m^x}^2 -{m^y}^2}} +\omega_m^2-\frac{2^{3/2} m^x m^p \lambda \omega_z}{\sqrt{\frac{1}{4} -{m^x}^2 -{m^y}^2}} + \omega_z^2) \chi^2 \\
    &+(\frac{m^y m^p \kappa^2 \lambda}{\sqrt{2} \sqrt{\frac{1}{4} -{m^x}^2 -{m^y}^2}} +\frac{2^{3/2} m^y m^p \lambda \omega_m^2}{\sqrt{\frac{1}{4} -{m^x}^2 -{m^y}^2}}-\frac{2^{3/2} m^x m^p \kappa \lambda \omega_z}{\sqrt{\frac{1}{4} -{m^x}^2 -{m^y}^2}}+\kappa \omega_z^2) \chi \\
    &-\frac{m^x m^p \kappa^2 \lambda \omega_z}{\sqrt{2} \sqrt{\frac{1}{4} -{m^x}^2 -{m^y}^2}}-8\sqrt{\frac{1}{4} -{m^x}^2 -{m^y}^2} \lambda^2 \omega_m \omega_z - \frac{2^{3/2} m^x m^p \lambda \omega_m^2 \omega_z }{\sqrt{\frac{1}{4} -{m^x}^2 -{m^y}^2}}+ \frac{\kappa^2 \omega_z^2}{4} \\
    =& a_0 \chi^4 +a_1 \chi^3 +a_2 \chi^2 + a_3 \chi +a_4 \overset{!}{=}0.
\end{align*}
Then,  Hurwitz' theorem \cite{Hurwitz,Gantmacher} states that all roots of this polynomial have negative real parts if and only if the inequalities 
\begin{align*}
    a_1&>0\\
    a_1 a_2 -a_0 a_3&>0\\
    (a_1 a_2 -a_0 a_3) a_3 -a_1^2 a_4 &>0\\
    a_4&>0
\end{align*}
hold. Employing this theorem we see that for our choice of parameters $\omega_z=4$, $\omega_m=1$, $\kappa=1$ all roots of the characteristic polynomial have a negative real part if and only if $m^z=-1/2$ and $0<\lambda<\lambda_c$ in the trivial stationary solution and $m^z=-\lambda_c^2/(2\lambda^2)$ and $\lambda>\lambda_c$ in the non-trivial solutions. Then the respective stationary solutions are asymptotically stable \cite{NonlinearSystems}. Moreover, numerical evidence shows that for $\lambda=\lambda_c$ small perturbations of the trivial solution (coinciding at this point with the non-trivial solutions) as initial conditions of the dynamics still drive the system to the trivial solution.\\
We proceed with the case $\gamma>0$. Here the Jacobian is 
\begin{align*}
J(\vec{m})= \begin{pmatrix}
 -\frac{\gamma}{2} & -\omega_z  & 0 & 0 & 0 \\
 \omega_z  & -\frac{\gamma}{2} & -2^{3/2} \lambda m^p & 0 & -2^{3/2} \lambda m^z\\
 0 & 2^{3/2} \lambda m^p & -\gamma &0 & 2^{3/2} \lambda m^y \\
  2^{3/2} \lambda & 0 & 0 & -\frac{\kappa}{2} & \omega_m \\
 0 & 0 & 0 & -\omega_m & -\frac{\kappa}{2}
\end{pmatrix}
\end{align*}
and the characteristic polynomial 
\begin{align*}
    \det( \mathrm{J}(\vec{m})- \mathds{1} \chi)=&-\chi^5-(2 \gamma+\kappa)\chi^4-(\frac{5}{4} \gamma^2 +2\gamma \kappa+\frac{1}{4} \kappa^2+8 {m^p}^2 \lambda^2+\omega_m^2+\omega_z^2)\chi^3\\
    &-(\frac{1}{4} \gamma^3+\frac{5}{4} \gamma^2 \kappa +\frac{1}{2} \gamma \kappa^2 +4 {m^p}^2 \gamma \lambda^2 + 8 {m^p}^2 \kappa \lambda^2 +2 \gamma \omega_m^2 + \gamma \omega_z^2 + \kappa \omega_z^2)\chi^2 \\
    &-(\frac{1}{4} \gamma^3 \kappa +\frac{5}{16} \gamma^2 \kappa^2 +4 {m^p}^2 \gamma \kappa \lambda^2+2 {m^p}^2 \kappa^2 \lambda^2+\frac{5}{4} \gamma^2 \omega_m^2+8 {m^p}^2 \lambda^2 \omega_m^2 + 8 m^z \lambda^2 \omega_m \omega_z +\gamma \kappa \omega_z^2 \\
    &+ \frac{1}{4} \kappa^2 \omega_z^2 +\omega_m^2 \omega_z^2)\chi -\frac{1}{16} \gamma^3 \kappa^2 -{m^p}^2 \gamma \kappa^2 \lambda^2 - \frac{1}{4} \gamma^3 \omega_m^2 - 4 {m^p}^2 \gamma \lambda^2 \omega_m^2 - 8 m^z \gamma \lambda^2 \omega_m \omega_z \\
    &- 2^{9/2} m^y m^p \lambda^3 \omega_m \omega_z - \frac{1}{4} \gamma \kappa^2 \omega_z^2 - \gamma \omega_m^2 \omega_z^2=-b_0 \chi^5-b_1 \chi^4 -b_2 \chi^3 -b_3 \chi^2 - b_4 \chi -b_5 \overset{!}{=}0.
\end{align*}
All roots of this degree $5$ polynomial have negative real parts (which is a necessary condition for stability of the solution)  \cite{Hurwitz,Gantmacher} if and only if the inequalities 
\begin{align*}
    b_1&>0\\
    b_1 b_2 -b_0 b_3&>0\\
    (b_1 b_2 -b_0 b_3) b_3 -b_1^2 b_4 +b_0 b_1 b_5&>0\\
    ((b_1 b_2 -b_0 b_3) b_3 -b_1^2 b_4 +b_0 b_1 b_5)b_4 +(b_2 b_3 +b_1 b_4) b_0 b_5 -b_0^2 b_5^2 -b_1 b_2^2 b_5&>0\\
    b_5&>0
\end{align*}
hold.
For $\omega_z=4$, $\omega_m=1$, $\kappa=1$ the trivial solution is asymptotically stable for $0\leq \lambda< \lambda_c$. On the other hand, for $\lambda>\lambda_c$ the non-trivial solutions are stable. Also in this case there is numerical evidence that for $\lambda=\lambda_c$ the stationary solution is approached eventually.

\section{Time-evolution of the covariance matrix}
In this section, we give the derivation of the dynamics of the covariance matrix reported in the main text. We will first derive the dynamics of the covariance matrix in the original frame and then show how this is modified when considering the emergent normal mode for the collective spin fluctuations. Finally, we discuss how the asymptotic covariance matrix can be found from the dynamical equation. 

\subsection{Derivation of the dynamics for fluctuations}
We start considering the original fluctuation operators that we collect in the following vector
\begin{align}
(F_N^{\alpha})_{\alpha=x,y,z,q,p} & =
					\begin{pmatrix} 
					(S^x - \langle S^x \rangle )/\sqrt{N}\\
					(S^y - \langle S^y \rangle )/\sqrt{N}\\
					(S^z - \langle S^z \rangle )/\sqrt{N}\\
					q- \langle q \rangle\\
					p - \langle p \rangle
					\end{pmatrix}.
					\label{fluc_SM}
\end{align}
Defining the two-point functions $C_N^{\alpha \beta} := \langle F_N^{\alpha}F_N^{\beta} \rangle$, the entries of the covariance matrix in the original frame $\Sigma$ are given by 
\begin{align} \label{CMCorr}
\Sigma_N^{\alpha \beta}
		=\frac{1}{2} \langle \{ F_N^{\alpha},F_N^{\beta} \} \rangle
		=\frac{1}{2} (\langle F_N^{\alpha}F_N^{\beta} \rangle + \langle F_N^{\beta}F_N^{\alpha} \rangle)=\frac{C_N^{\alpha \beta}+{C_N^T}^{\alpha \beta}}{2}.
\end{align}
In general, the operators of the form $F_N^{\alpha}F_N^{\beta}$ possess an explicit time-dependence through expectation values over the state contained in the definition of fluctuations in Eq.~\eqref{fluc_SM}. Taking the total time-derivative of the two-point functions we thus get 
\begin{align*}
   \frac{d}{dt} \langle F^{\alpha}_{N} F^{\beta}_{N} \rangle=  \langle \mathcal{L}_N[ F^{\alpha}_{N} F^{\beta}_{N}] \rangle +\frac{d}{dt}(F^{\alpha}_{N}) \langle  F^{\beta}_{N} \rangle +\frac{d}{dt}(F^{\beta}_{N}) \langle F^{\alpha}_{N}  \rangle= \langle \mathcal{L}_N[ F^{\alpha}_{N} F^{\beta}_{N}] \rangle.
\end{align*}
Here, we used that $\langle F^{\alpha}_{N} \rangle =0$ by definition and that $d/dt F^{\alpha}_{N}$ is a scalar quantity. We thus have that  
\begin{align} \label{Corr}
\dot{C}_N^{\alpha \beta} 
		=\langle iF_N^{\alpha}[H_N^D, F_N^{\beta}] \rangle +\langle i[H_N^D, F_N^{\alpha}]F_N^{\beta} \rangle + \langle \mathcal{D}_N^{\kappa}[F_N^{\alpha}F_N^{\beta}]   \rangle + \langle \mathcal{D}_N^{\gamma}[F_N^{\alpha}F_N^{\beta}]   \rangle 
\end{align}
Thus the problem is, like in the mean-field analysis, separated into three parts. First we want to evaluate the first two terms. To this end, we consider the commutator of the Dicke Hamiltonian and the fluctuation vector components and Eq. (\ref{H_N^Dcommutator}) whence
\begin{align*}
    i[H_N^D,F_N^{\alpha}]=&(\delta^{\alpha x}+\delta^{\alpha y}+\delta^{\alpha z})( \frac{-\omega_z}{\sqrt{N}} \epsilon^{z \alpha \gamma} S^{\gamma} - \frac{2^{3/2}}{\sqrt{N}} \lambda m_N^p \epsilon^{x \alpha \gamma} S^{\gamma})+\delta^{\alpha q}(\omega_m p  + \frac{2^{3/2} \lambda}{\sqrt{N}} S^x)-\delta^{\alpha p}\omega_m q .
\end{align*}
Exploiting the fact that $\langle F_N^{\alpha} \rangle=0$, we can write 
\begin{align*}
    \langle i[H_N^D, F_N^{\alpha}]F_N^{\beta} \rangle=\langle i[H_N^D, F_N^{\alpha}]F_N^{\beta} \rangle-\langle i[H_N^D, F_N^{\alpha}]\rangle \langle F_N^{\beta} \rangle \overset{N \gg 1}{=}&(\delta^{\alpha x}+\delta^{\alpha y}+\delta^{\alpha z})( -\omega_z \epsilon^{z \alpha \gamma} \langle F^{\gamma} F^{\beta} \rangle - 2^{3/2} \lambda \epsilon^{x \alpha \gamma}  m^{\gamma}  \langle   F^{p} F^{\beta}\rangle \\-& 2^{3/2} \lambda \epsilon^{x \alpha \gamma}  m^p  \langle  F^{\gamma}   F^{\beta}\rangle)
    +\delta^{\alpha q}(\omega_m \langle F^{p} F^{\beta} \rangle  + 2^{3/2} \lambda \langle F^{x} F^{\beta} \rangle)\\-&\delta^{\alpha p}\omega_m \langle F^{q} F^{\beta} \rangle \\
    =&: A^{\alpha \gamma} C^{\gamma \beta}=(AC)^{\alpha \beta}
\end{align*}
with 
\begin{align*}
A 
= \begin{pmatrix}
 0 & -\omega_z & 0 & 0 &  0\\
 \omega_z & 0 & -2^{3/2} \lambda  m^p  & 0 & -2^{3/2} \lambda  m^z \\
 0 & 2^{3/2} \lambda  m^p   & 0 & 0 & 2^{3/2} \lambda  m^y \\
 2^{3/2} \lambda & 0 & 0 & 0 & \omega_m \\
 0 & 0 & 0 & -\omega_m & 0\\
\end{pmatrix}.
\end{align*}
In the above calculation we have used that $m_N^\alpha\to m^\alpha$, multiple of the identity, in the thermodynamic limit. 

Analogously, we can calculate 
\begin{align*}
    \langle i F_N^{\alpha}[H_N^D,F_N^{\beta}] \rangle\overset{N \gg 1}{=}&(\delta^{\beta x}+\delta^{\beta y}+\delta^{\beta z})( -\omega_z \epsilon^{z \beta \gamma} \langle F^{\alpha} F^{\gamma}  \rangle - 2^{3/2} \lambda \epsilon^{x \beta \gamma}  m^{\gamma}  \langle F^{\alpha}   F^{p} \rangle- 2^{3/2} \lambda \epsilon^{x \beta \gamma}  m^p  \langle F^{\alpha}  F^{\gamma}  \rangle)\\
    &+\delta^{\beta q}(\omega_m \langle F^{\alpha} F^{p} \rangle  + 2^{3/2} \lambda \langle F^{\alpha} F^{x} \rangle)-\delta^{\beta p}\omega_m \langle F^{\alpha} F^{q} \rangle \\
    =:& C^{\alpha \gamma} B^{\gamma \beta}=(CB)^{\alpha \beta}
\end{align*}
with 
\begin{align*}
B 
= \begin{pmatrix}
 0 & \omega_z & 0 & 2^{3/2} \lambda &  0\\
 -\omega_z & 0 & 2^{3/2} \lambda  m^p  & 0 & 0\\
 0 & -2^{3/2} \lambda  m^p  & 0 & 0 & 0\\
 0  & 0 & 0 & 0 & -\omega_m \\
 0 & -2^{3/2} \lambda  m^z  & 2^{3/2} \lambda  m^y  & \omega_m & 0\\
\end{pmatrix}=A^T.
\end{align*}
For the remaining two parts of $\dot{C}_N$ we expand 
\begin{align*}
    \frac{1}{2} ([A^{\dagger},F_N^{\alpha}F_N^{\beta}]A +A^{\dagger} [F_N^{\alpha}F_N^{\beta},A])=& F_N^{\alpha}(\frac{1}{2} ([A^{\dagger},F_N^{\beta}]A +A^{\dagger} [F_N^{\beta},A])) 
    + (\frac{1}{2} ([A^{\dagger},F_N^{\alpha}]A +A^{\dagger} [F_N^{\alpha},A]))F_N^{\beta} 
    +[A^{\dagger},F_N^{\alpha}][F_N^{\beta},A]
\end{align*}
to achieve 
\begin{align*}
    \mathcal{D}_N^{\kappa}[F_N^{\alpha}F_N^{\beta}]= F_N^{\alpha} \mathcal{D}_N^{\kappa}[F_N^{\beta}]+ \mathcal{D}_N^{\kappa}[F_N^{\alpha}]F_N^{\beta}+ \frac{N\kappa}{2} [m_N^p+im_N^q,F_N^{\alpha}][F_N^{\beta},m_N^p-im_N^q]
\end{align*}
and 
\begin{align} \label{GDissFluct}
    \mathcal{D}_N^{\gamma}[F_N^{\alpha}F_N^{\beta}]= F_N^{\alpha} \mathcal{D}_N^{\gamma}[F_N^{\beta}]+ \mathcal{D}_N^{\gamma}[F_N^{\alpha}]F_N^{\beta}+ \gamma \sum_{k=1}^N [\sigma_k^+,F_N^{\alpha}][F_N^{\beta},\sigma_k^-].
\end{align}
Focusing on $\mathcal{D}_N^{\kappa}$, we find for the last term on the right-hand side 
\begin{align*}
    \frac{\kappa}{2} [p+iq,F_N^{\alpha}][F_N^{\beta},p-iq] &=\frac{\kappa}{2} (-s_N^{ \alpha p} s_N^{p \beta }-is_N^{ \alpha q}s_N^{p \beta }+is_N^{ \alpha p}s_N^{ q \beta }-s_N^{ \alpha q} s_N^{q \beta })=:-s_N^{\alpha \gamma} D'^{\gamma \delta} s_N^{\delta \beta}=(-s_ND's_N)^{\alpha \beta}
\end{align*}
where the \textit{symplectic matrix} $s_N$ is given by the commutation relations of the fluctuation operators $s_N^{\alpha \beta}=-i[F_N^{\alpha},F_N^{\beta}]$ and explicitly 
\begin{align*}
s_N
\overset{N \gg 1}{=} \begin{pmatrix}
 0 &   m^z   & -  m^y   & 0 &  0\\
 -  m^z  & 0 &  m^x  & 0 & 0\\
  m^y  & -  m^x  & 0 & 0 & 0\\
 0 & 0 & 0 & 0 & 1\\
 0 & 0 & 0 & -1 & 0
\end{pmatrix}=s.
\end{align*}
Furthermore 
\begin{align*}
D'
=\frac{\kappa}{2}\begin{pmatrix}
 0 &  0  & 0  & 0 &  0\\
 0  & 0 & 0 & 0 & 0\\
 0 & 0  & 0 & 0 & 0\\
 0 & 0 & 0 & 1 & i\\
 0 & 0 & 0 & -i & 1
\end{pmatrix}.
\end{align*}
The single-fluctuation $\kappa$-dissipator reads 
\begin{align*}
    \mathcal{D}_N^{\kappa}[F_N^{\alpha}] =& \frac{ \kappa}{4} (is_N^{p \alpha}p+s_N^{p \alpha}q-s_N^{q \alpha}p+is_N^{q \alpha}q +ip s_N^{\alpha p}-q s_N^{\alpha p}+p s_N^{\alpha q}+iq s_N^{\alpha q})
\end{align*}
and therefore 
\begin{align*}
    \langle \mathcal{D}_N^{\kappa}[F_N^{\alpha}]F_N^{\beta} \rangle \overset{N \gg 1}{=} \frac{ \kappa}{2} ( s^{\alpha q}\langle F^{p} F^{\beta} \rangle -s^{\alpha p}\langle F^{q} F^{\beta} \rangle)=:s^{\alpha \gamma} E^{\gamma \delta} C^{\delta \beta}=(sEC)^{\alpha \beta}.
\end{align*}
Analogously 
\begin{align*}
    \langle F_N^{\alpha}\mathcal{D}_N^{\kappa}[F_N^{\beta}] \rangle \overset{N\gg 1}{=} \frac{ \kappa}{2} ( -\langle F^{\alpha} F^{p}  \rangle s^{ q \beta} +\langle F^{\alpha} F^{q}  \rangle s^{ p \beta})=: C^{\alpha \gamma} E'^{\gamma \delta} s^{\delta \beta}=(CE's)^{\alpha \beta}
\end{align*}
with 
\begin{align*}
E
=\frac{\kappa}{2}\begin{pmatrix}
 0 &  0  & 0  & 0 &  0\\
 0  & 0 & 0 & 0 & 0\\
 0 & 0  & 0 & 0 & 0\\
 0 & 0 & 0 & 0 & 1\\
 0 & 0 & 0 & -1 & 0
\end{pmatrix}=E'.
\end{align*}
Collecting intermediately all the results concerning $\mathcal{D}_N^{\kappa}$, we see 
\begin{align*}
    \langle \mathcal{D}_N^{\kappa}[F_N^{\alpha}F_N^{\beta}] \rangle \overset{N \gg 1}{=}(CEs+sEC-sD's)^{\alpha \beta}.
\end{align*}
Proceeding with $\mathcal{D}_N^{\gamma}$, we have for the single-fluctuation dissipator 
\begin{align*}
\mathcal{D}_N^{\gamma}[F_N^{\alpha}]&= \frac{\gamma}{2 \sqrt{N}} \sum_{k=1}^N(\delta^{x \alpha}(\sigma_k^z \sigma_k^- +\sigma_k^+ \sigma_k^z)+\delta^{y \alpha}(i \sigma_k^z \sigma_k^- -i\sigma_k^+\sigma_k^z)+\delta^{z \alpha}(-\sigma_k^+ \sigma_k^- -\sigma_k^+\sigma_k^-))\\
&= \frac{\gamma}{2} (\delta^{x \alpha}(-\frac{S^x}{\sqrt{N}})+\delta^{y \alpha}(-\frac{S^y}{\sqrt{N}})+\delta^{z \alpha}(-\sqrt{N}\mathds{1}-2\frac{S^z}{\sqrt{N}}))
\end{align*}
and 
\begin{align*}
\langle \mathcal{D}_N^{\gamma}[F_N^{\alpha}] F_N^{\beta} \rangle & \overset{N \gg 1}{=}  \frac{\gamma}{2} (\delta^{x \alpha}(-\langle F^x F^{\beta} \rangle)+\delta^{y \alpha}(-\langle F^y F^{\beta} \rangle )+\delta^{z \alpha}(-2 \langle F^z F^{\beta} \rangle)) \\
&=: Q^{\alpha \gamma}C^{\gamma \beta}=(QC)^{\alpha \beta}
\end{align*}
with 
\begin{align*}
Q= \begin{pmatrix}
 -\frac{\gamma}{2} & 0 & 0 & 0 &  0\\
 0 &  -\frac{\gamma}{2} & 0 & 0 & 0\\
 0 & 0 & -\gamma & 0 & 0\\
 0 & 0 & 0 & 0 & 0\\
 0 & 0 & 0 & 0 & 0
\end{pmatrix}.
\end{align*}
Also it is 
\begin{align*}
\langle F_N^{\alpha} \mathcal{D}_N^{\gamma}[F_N^{\beta}]  \rangle & \overset{N \gg 1}{=}  \frac{\gamma}{2} (\delta^{x \beta}(-\langle F^{\alpha} F^x  \rangle)+\delta^{y \beta}(-\langle F^{\alpha} F^y \rangle )+\delta^{z \beta}(-2 \langle F^{\alpha} F^z \rangle)) \\
&= C^{\alpha \gamma}Q^{\gamma \beta}=(CQ)^{\alpha \beta}.
\end{align*}
For the last term in Eq. (\ref{GDissFluct}) we get 
\begin{align*}
 \gamma \sum_{k=1}^N [\sigma_k^+,F_N^{\alpha}][F_N^{\beta},\sigma_k^-]  &=  \frac{\gamma}{N} \sum_{k=1}^N (\delta^{x \alpha}\delta^{x \beta}\frac{\mathds{1}_k}{4}+\delta^{x \alpha}\delta^{y \beta} \frac{-i\mathds{1}_k}{4}+\delta^{x \alpha}\delta^{z \beta} (-\sigma_k^z \sigma_k^-)\\
 &  +\delta^{y \alpha}\delta^{x \beta}\frac{i\mathds{1}_k}{4}+\delta^{y \alpha}\delta^{y \beta}\frac{\mathds{1}_k}{4}+\delta^{y \alpha}\delta^{z \beta} (-i \sigma_k^z \sigma_k^-)\\
 &+\delta^{z \alpha}\delta^{x \beta} (-\sigma_k^+ \sigma_k^z)+\delta^{z \alpha}\delta^{y \beta} i\sigma_k^+ \sigma_k^z +\delta^{z \alpha}\delta^{z \beta} \sigma_k^+ \sigma_k^-)
\end{align*}
and by means of $\sigma_k^{\rho}\sigma_k^{\nu}=\delta^{\rho \nu}\frac{\mathds{1}_k}{4}+i \epsilon^{\rho \nu \mu} \frac{\sigma_k^{\mu}}{2}$,
\begin{align*}
\langle \gamma \sum_{k=1}^N [\sigma_k^+,F_N^{\alpha}][F_N^{\beta},\sigma_k^-] \rangle  &\overset{N \gg 1}{=}  \gamma (\delta^{x \alpha}\delta^{x \beta}\frac{1}{4}+\delta^{x \alpha}\delta^{y \beta} \frac{-i}{4}+\delta^{x \alpha}\delta^{z \beta} \frac{ m^x -i  m^y }{2}\\
 &  +\delta^{y \alpha}\delta^{x \beta}\frac{i}{4}+\delta^{y \alpha}\delta^{y \beta}\frac{1}{4}+\delta^{y \alpha}\delta^{z \beta} \frac{ m^y +i  m^x }{2}\\
 &+\delta^{z \alpha}\delta^{x \beta} \frac{ m^x +i  m^y  }{2}+\delta^{z \alpha}\delta^{y \beta} \frac{ m^y -i  m^x }{2} +\delta^{z \alpha}\delta^{z \beta} (\frac{1}{2} +  m^z )) \\
 &=:{Z'}^{\alpha \beta}.
\end{align*}
Here 
\begin{align*}
Z'=\gamma \begin{pmatrix}
 \frac{1}{4} & \frac{-i}{4} & \frac{ m^x -i  m^y }{2} & 0 &  0\\
 \frac{i}{4} &  \frac{1}{4} & \frac{ m^y +i  m^x }{2} & 0 & 0\\
 \frac{ m^x  +i  m^y }{2} & \frac{ m^y -i  m^x }{2} & \frac{1}{2}+ m^z  & 0 & 0\\
 0 & 0 & 0 & 0 & 0\\
 0 & 0 & 0 & 0 & 0
\end{pmatrix}.
\end{align*}
Now also collecting the results concerning $\mathcal{D}_N^{\gamma}$ gives 
\begin{align*}
    \langle \mathcal{D}_N^{\gamma}[F_N^{\alpha}F_N^{\beta}] \rangle \overset{N \gg 1}{=} (CQ+QC+Z')^{\alpha \beta}
\end{align*}
and we conclude for Eq. (\ref{Corr}) in the thermodynamic limit
\begin{align*}
    \dot{C}=CA^T+AC+CEs+sEC-sD's+CQ+QC+Z'.
\end{align*}
Considering then Eq.~\eqref{CMCorr}, we finally get the  differential equation for the covariance matrix 
\begin{align*}
    \dot{\Sigma}=& \Sigma(A^T+Es+Q)+(A+ s E+Q)\Sigma- \frac{s D' s +s^TD'^Ts^T}{2} + \frac{Z'+Z'^T}{2}\\
    =& \Sigma G^T+G\Sigma -s D s+Z
\end{align*}
where we defined $G:=A+sE+Q$ and 
\begin{align*}
Z:=\gamma \begin{pmatrix}
 \frac{1}{4} & 0 & \frac{ m^x }{2} & 0 &  0\\
 0 &  \frac{1}{4} & \frac{ m^y }{2} & 0 & 0\\
 \frac{ m^x }{2} & \frac{ m^y }{2} & \frac{1}{2}+ m^z  & 0 & 0\\
 0 & 0 & 0 & 0 & 0\\
 0 & 0 & 0 & 0 & 0
\end{pmatrix}, 
\qquad
D:=\frac{\kappa}{2}\begin{pmatrix}
 0 &  0  & 0  & 0 &  0\\
 0  & 0 & 0 & 0 & 0\\
 0 & 0  & 0 & 0 & 0\\
 0 & 0 & 0 & 1 & 0\\
 0 & 0 & 0 & 0 & 1
\end{pmatrix}.
\end{align*}

We note that the differential equation for the covariance matrix involves, in general, time-dependent matrices $G,s$ and $Z$. These may indeed be time-dependent through the time-dependence of the mean-field operators which appear in their matrix elements. However, for the purpose of this work, the matrices $G,s$ and $Z$ are time-independent since we investigate here the behavior of fluctuations when the state of the system is already stationary with respect to the mean-field observables. 

\subsection{Emergent normal mode}
We transform the fluctuation vector as 
\begin{align*}
&F^{\alpha} \rightarrow r^{\alpha}=J R_{\theta,\varphi} F^{\alpha} \quad \textrm{where} \quad 
J:=\begin{pmatrix}
 1/\sqrt{| \vec{m}  |} &  0  & 0  & 0 &  0\\
 0  &  1/\sqrt{| \vec{m}  |} & 0 & 0 & 0\\
 0 & 0  & 1 & 0 & 0\\
 0 & 0 & 0 & 1 & 0\\
 0 & 0 & 0 & 0 & 1
\end{pmatrix}
\end{align*}
\textrm{and}
\begin{align*}
R_{\theta,\varphi}&:=
\begin{pmatrix}
 \cos \theta & 0 & - \sin \theta & 0 & 0\\
 0 & 1 & 0 & 0 & 0\\
 \sin \theta & 0 & \cos \theta  & 0 & 0 \\
 0 & 0 & 0 & 1 & 0 \\
 0 & 0 & 0 & 0 & 1
\end{pmatrix}
\begin{pmatrix}
 \cos \varphi & \sin \varphi & 0 & 0 & 0\\
 - \sin \varphi & \cos \varphi & 0 & 0 & 0\\
 0 & 0 & 1 & 0 & 0 \\
 0 & 0 & 0 & 1 & 0 \\
 0 & 0 & 0 & 0 & 1
\end{pmatrix}  \\
&=\begin{pmatrix}
 \cos \theta \cos \varphi &  \cos \theta \sin \varphi & - \sin \theta & 0 & 0\\
 -\sin \varphi & \cos \varphi & 0 & 0 & 0\\
 \sin \theta \cos \varphi & \sin \theta \sin \varphi & \cos \theta & 0 & 0 \\
 0 & 0 & 0 & 1 & 0 \\
 0 & 0 & 0 & 0 & 1
\end{pmatrix}.
\end{align*}
The matrix $R_{\theta,\varphi}$ represents a rotation of the spin-part of the fluctuation vector by an angle $-\varphi$ around the $z$-axis and by an angle $-\theta$ around the $y$-axis. The angles are 
\begin{align*}
\theta = \arccos ( \frac{ m^z }{| \vec{m}  |})
\quad \textrm{and} \quad 
\varphi = 
\begin{cases}
2 \pi - \arccos (\frac{ m^x }{| \vec{m}  | \sin \theta }) &,  m^y  < 0 \\
0 & ,  m^y  = 0, \;  m^x  = 0 \\
\arccos (\frac{ m^x }{| \vec{m}  | \sin \theta }) &  \textrm{otherwise}.
\end{cases}
\end{align*}
The $J$-matrix realizes a rescaling of the two remaining non-classical collective spin degrees of freedom, obtained after rotating, such that indeed 
\begin{align*}
\tilde{s}^{\alpha \beta} & =-i  [ (JR_{\theta,\varphi} F)^{\alpha} , (JR_{\theta,\varphi} F)^{\beta} ] =  -i  [ J^{\alpha \nu} R_{\theta,\varphi}^{\nu \epsilon} F^{\epsilon} , J^{\beta \delta} R_{\theta,\varphi}^{\delta \gamma} F^{\gamma} ]\\
&=    J^{\alpha \nu} R_{\theta,\varphi}^{\nu \epsilon}s^{\epsilon \gamma}  ( R_{\theta,\varphi}^T J^T)^{\gamma \beta}= (JR_{\theta,\varphi} s R_{\theta,\varphi}^T J)^{\alpha \beta}
\end{align*}
and 
\begin{align*}
\tilde{s}=
\begin{pmatrix}
0 & 1 & 0 & 0 & 0 \\
-1 & 0 & 0 & 0 & 0 \\
0 & 0 & 0 & 0 & 0 \\
0 & 0 & 0 & 0 & 1 \\
0 & 0 & 0 & -1 & 0 
\end{pmatrix}.
\end{align*}
Similarly the covariance matrix  transforms as $\tilde{\Sigma}=JR_{\theta,\varphi}\Sigma R_{\theta,\varphi}^T J$ and the time-evolution is now given by the differential equation 
\begin{align*}
\dot{\tilde{\Sigma}}  = & J R_{\theta,\varphi} (Z-sDs + \Sigma G^T + G \Sigma) R_{\theta,\varphi}^T J\\
 = & JR_{\theta,\varphi}Z R_{\theta,\varphi}^T J -\tilde{s}(J^{-1} R_{\theta,\varphi} D R_{\theta,\varphi}^T J^{-1} ) \tilde{s} + \tilde{\Sigma} (J R_{\theta,\varphi} G R_{\theta,\varphi}^T J^{-1})^T + (J R_{\theta,\varphi} G R_{\theta,\varphi}^T J^{-1}) \tilde{\Sigma}\\
 = & \tilde{Z}-\tilde{s} \tilde{D} \tilde{s} + \tilde{\Sigma} \tilde{G}^T + \tilde{G} \tilde{\Sigma}\, ,
\end{align*}
where for the sake of clarity we have 
$$
\tilde{G}=JR_{\theta,\varphi}GR_{\theta,\varphi}^TJ^{-1}\, ,\quad \tilde{Z}=JR_{\theta,\varphi}ZR_{\theta,\varphi}^TJ\, ,\quad \textrm{and} \quad \tilde{D}=J^{-1} R_{\theta,\varphi} D R_{\theta,\varphi}^T J^{-1} =D\, .
$$
We stress here again that the considered initial state for the system is stationary with respect to the mean-field observables so that also $R_{\theta,\varphi}$ and $J$ are time-independent. 

\subsection{Stationary covariance matrix}
In this section we show how the stationary covariance matrix can be obtained through a vectorization procedure. We will refer to an odd-dimensional square matrix as in ``cross" form if its middle row and column consist only of zeros. An even-dimensional square matrix arising from an odd-dimensional one $M$ by deleting the middle row and column is said to be in ``reduced" form and we denote it by $M_{\mathrm{red}}$.\\
We start with the differential equation for the covariance matrix from the last section
\begin{align}
    \dot{\tilde{\Sigma}}(t)  = \tilde{\Sigma}(t) \tilde{G}^T + \tilde{G} \tilde{\Sigma}(t) +  \tilde{W} \label{Sylvester}
\end{align}
where we defined $\tilde{W}=\tilde{Z}-\tilde{s} \tilde{D} \tilde{s}$. We focus on parameters chosen for Fig.~\ref{Fig1} in the main text, i.~e.~$\omega_z=4$, $\omega_m=1$, $\kappa=1$. Let $I\subseteq \mathbb{R}$ be an open interval. Here $t\in I$ and $t>t_0 \in I$. At $t_0$ it is assumed that the quantum state is such that 
\begin{align*}
    \vec{m}(t_0)=\begin{cases}
    \vec{m}_{\mathrm{sub}} & \mathrm{for} \ \lambda\in[0,\lambda_c] \\
    \vec{m}_{\mathrm{sup}} & \mathrm{for} \ \lambda\in(\lambda_c,\infty)\, ,
    \end{cases}
\end{align*}
where $\vec{m}_{\mathrm{sub}}$ is the vector containing the stable solution of the mean-field equations in the normal phase, while $\vec{m}_{\mathrm{sup}}$ is the vector containing the solution of the mean-field equations in the superradiant phase. 

The cases $\gamma=0$ and $\gamma>0$ are treated separately and we first focus on $\gamma>0$. The task is to find the stationary covariance matrix $\tilde{\Sigma}_\infty$ which is such that $\dot{\tilde{\Sigma}}_\infty=0$. The matrix equation to be solved, given by  
\begin{align*}
       \tilde{\Sigma}_\infty \tilde{G}^T + \tilde{G} \tilde{\Sigma}_\infty =- \tilde{W}\, ,
\end{align*}
is equivalent \cite{MatrixAnalysis} to finding the $25$ unknowns of the following linear system of $25$ equations 
\begin{align}
       ( \tilde{G} \otimes \mathds{1}_5 + \mathds{1}_5 \otimes \tilde{G})  vec(\tilde{\Sigma}_\infty)= vec( -\tilde{W}) \label{LinearEquation}
\end{align}
where $\otimes$ is the Kronecker product and the operation $vec(\cdot)$ arranges the entries of a matrix columnwise in a vector top down. Eq.~\eqref{LinearEquation} has a unique solution  if and only if $\tilde{G} \otimes \mathds{1}_5 + \mathds{1}_5 \otimes \tilde{G}$ is invertible. The solution is, in vectorized form, given by 
\begin{align*}
vec(\tilde{\Sigma}_{\infty})= ( \tilde{G} \otimes \mathds{1}_5 + \mathds{1}_5 \otimes \tilde{G})^{-1} vec( -\tilde{W}).
\end{align*}
Equivalently to invertibility, we want to prove that any eigenvalue of $\tilde{G} \otimes \mathds{1}_5 + \mathds{1}_5 \otimes \tilde{G}$ is nonzero. If the spectrum of $\tilde{G}$ is $\sigma(\tilde{G})=\{\mu_1,\mu_2,\mu_3,\mu_4,\mu_5\}$ then the set of these eigenvalues is $\sigma(\tilde{G} \otimes \mathds{1}_5 + \mathds{1}_5 \otimes \tilde{G})=\{ \mu_i +\mu_j | i=1,...,5, j=1,...,5\}$. Thus, any eigenvalue is nonzero if $\sigma(\tilde{G}) \cap \sigma(-\tilde{G})=\emptyset$, i.e. if no element of $\sigma(-\tilde{G})$ can be obtained by a point reflection of an element of $\sigma(\tilde{G})$ at the origin of the complex plane. Using again Hurwitz' theorem it can be proven that for $\lambda\neq \lambda_c$ all eigenvalues of $\tilde{G}$ lie in the open left half-plane. Consequently the matrix $\tilde{G} \otimes \mathds{1}_5 + \mathds{1}_5 \otimes \tilde{G}$ is invertible if $\lambda\neq \lambda_c$. 

In the $\gamma=0$ case one cannot proceed the same way. In this setting, we focus on initial covariance matrices $\tilde{\Sigma}(t_0)$ that are in cross form (see the definition at the beginning of this subsection). The matrix $\tilde{Z}$ is the zero matrix and the differential equation \eqref{Sylvester} reduces to 
\begin{align*}
    \dot{\tilde{\Sigma}}(t)  = -\tilde{s} \tilde{D} \tilde{s} + \tilde{\Sigma}(t) \tilde{G}^T + \tilde{G} \tilde{\Sigma}(t).
\end{align*}
Known as the differential Sylvester equation \cite{LyapEqu}, it has the unique solution 
\begin{align}
\tilde{\Sigma}(t)=e^{\tilde{G}(t-t_0)}\tilde{\Sigma}(t_0)e^{\tilde{G}^T(t-t_0)}-\int^t_{t_0} e^{\tilde{G}(t-s)}\tilde{s} \tilde{D} \label{CMSol} \tilde{s}e^{\tilde{G}^T(t-s)} ds.
\end{align}
We note that 
\begin{align*}
\tilde{s}=
\begin{pmatrix}
0 & 1 & 0 & 0 & 0 \\
-1 & 0 & 0 & 0 & 0 \\
0 & 0 & 0 & 0 & 0 \\
0 & 0 & 0 & 0 & 1 \\
0 & 0 & 0 & -1 & 0 
\end{pmatrix} \ , \
\tilde{G}
=\begin{pmatrix}
0 & -\frac{2^{3/2}  m^p  \lambda  m^x  +  m^z  \omega_z}{| \vec{m} |} & 0 & 0 & 0 \\
\frac{2^{3/2}  m^p  \lambda  m^x  +  m^z  \omega_z}{| \vec{m} |} & 0 & 0 & 0 & \frac{-\sgn( m^x ) 2^{3/2} \lambda  m^z }{\sqrt{| \vec{m} |}} \\
0 & 0 & 0& 0& 0 \\
\frac{\sgn( m^x )2^{3/2} \lambda  m^z }{\sqrt{| \vec{m} |}} & 0  & \frac{2^{3/2} \lambda  m^x }{| \vec{m} |} & -\frac{\kappa}{2} & \omega_c \\
0 & 0 & 0 & -\omega_c & -\frac{\kappa}{2}  
\end{pmatrix}
\end{align*}
and $\tilde{D}=D$. With $e^{\tilde{G}(t-t_0)}=\sum^{\infty}_{n=0}\frac{\tilde{G}^n}{n!} \cdot(t-t_0)^n$ it is
\begin{align*}
e^{\tilde{G}(t-t_0)}\tilde{\Sigma}(t_0)e^{\tilde{G}^T(t-t_0)}=\lim_{m,n \to \infty}\sum^{m}_{k=0}\sum^{n}_{l=0}\frac{(t-t_0)^k(t-t_0)^l}{k!l!}  \tilde{G}^k \tilde{\Sigma}(t_0) (\tilde{G}^T)^l.
\end{align*} 
This is in cross form  since $\tilde{\Sigma}(t_0)$ is in this form and thus $\tilde{G}^k \tilde{\Sigma}(t_0) (\tilde{G}^T)^l$  is in cross form, for all $k,l \in \mathbb{N}_0$. Similarly, since $\tilde{s} \tilde{D} \tilde{s}$ is in cross form, also $e^{\tilde{G}(t-s)}\tilde{s} \tilde{D} \tilde{s}e^{\tilde{G}^T(t-s)}$ is. It follows that the unique solution in Eq.~\eqref{CMSol} has cross form for all $t\in I$ with $t>t_0$. Therefore it remains to solve 
\begin{align*}
    \dot{\tilde{\Sigma}}_{\mathrm{red}}(t)  = \dot{\tilde{\Sigma}}^{\mathrm{t-m}}(t)=-\tilde{s}_{\mathrm{red}} \tilde{D}_{\mathrm{red}} \tilde{s}_{\mathrm{red}} + \tilde{\Sigma}^{\mathrm{t-m}}(t)  \tilde{G}^T_{\mathrm{red}} + \tilde{G}_{\mathrm{red}} \tilde{\Sigma}^{\mathrm{t-m}}(t) .
\end{align*}
We want to find the stationary covariance matrix $\dot{\tilde{\Sigma}}^{\mathrm{t-m}}_\infty(t) =0$. Solving the matrix equation 
\begin{align*}
  \tilde{\Sigma}^{\mathrm{t-m}}_\infty \tilde{G}^T_{\mathrm{red}} + \tilde{G}_{\mathrm{red}} \tilde{\Sigma}^{\mathrm{t-m}}_\infty= \tilde{s}_{\mathrm{red}} \tilde{D}_{\mathrm{red}} \tilde{s}_{\mathrm{red}}
\end{align*}
is equivalent to finding the $16$ unknowns of the linear system of $16$ equations 
\begin{align*}
  (\tilde{G}_{\mathrm{red}} \otimes \mathds{1}_4 + \mathds{1}_4 \otimes \tilde{G}_{\mathrm{red}} ) vec(\tilde{\Sigma}^{\mathrm{t-m}}_\infty) = vec(\tilde{s}_{\mathrm{red}} \tilde{D}_{\mathrm{red}} \tilde{s}_{\mathrm{red}}).
\end{align*}
With the same steps as above, we establish the invertibility of $\tilde{G}_{\mathrm{red}} \otimes \mathds{1}_4 + \mathds{1}_4 \otimes \tilde{G}_{\mathrm{red}} $ for $\lambda \not\in \{0,\lambda_c \}$ such that in this regime the unique stationary CM is given by 
\begin{align*}
vec(\tilde{\Sigma}^{\mathrm{t-m}}_\infty)= ( \tilde{G}_{\mathrm{red}} \otimes \mathds{1}_4 + \mathds{1}_4 \otimes \tilde{G}_{\mathrm{red}})^{-1} vec( \tilde{s}_{\mathrm{red}} \tilde{D}_{\mathrm{red}} \tilde{s}_{\mathrm{red}} ).
\end{align*}

\section{Quantum and classical correlations for two-mode bosonic Gaussian states}
Here we discuss important tools of Gaussian quantum information theory and define the correlation measures exploited in the main text. In addition, we report supplementary results.
\subsection{Measures of correlations for Gaussian bosonic systems}
The total correlations contained in a quantum state $\rho_{s,b}$ (the notation reflects the bipartition into a spin subsystem and a boson subsystem) can be measured by the \textit{quantum mutual information}
\begin{align*}
   I(\rho_{s,b})=S(\rho_s)+S(\rho_b)-S(\rho_{s,b}) 
\end{align*}
with the von Neumann entropy $S(\rho)=-\Tr(\rho \log \rho)$ and  $\rho_{s/b}$ being the reduced state for system $s/b$. Henceforth, $\log$ denotes the natural logarithm. The quantum mutual information can be written as sum of a purely classical part
\begin{align*}
\mathcal{J}(\rho_{s,b})=S(\rho_s)-\inf_{\{\Pi_i\}}\sum_i p_i S(\Tr_b(\rho_{s,b} \Pi_i)/p_i)\, ,
\end{align*}
and a quantum part $\mathcal{D}(\rho_{s,b})$ defined as the difference of $I$ and $\mathcal{J}$. In this decomposition of the quantum mutual information, $\mathcal{D}$ is called the \textit{quantum discord} and $\mathcal{J}$ the \textit{classical correlation} In the above equation, it is $p_i=\Tr_{s,b}(\rho_{s,b}\Pi_i)$ and the infimum is taken over all (Gaussian) positive operator-valued measures (POVMs) $\{\Pi_i\}$, $\sum_i\Pi_i=\mathds{1}$ on the boson system. At the covariance matrix level, considering the two-mode matrix 
\begin{align} \label{CMTwoMode}
\tilde{\Sigma}^{\mathrm{t-m}}_\infty
=  \begin{pmatrix}
\Gamma_s & \Gamma_c \\
\Gamma_c^T & \Gamma_b 
\end{pmatrix}
\end{align}
and explicitly carrying out the minimization leads for the definitions 
\begin{align*}
    A=\det (2\Gamma_s), \qquad B=\det (2\Gamma_b), \qquad C=\det (2\Gamma_c) \qquad \textrm{and} \qquad D=\det (2\Sigma^{\mathrm{t-m}}_\infty)
\end{align*}
to the closed expressions
\begin{align*}
\mathcal{J}(\tilde{\Sigma}^{\mathrm{t-m}}_\infty )=f(\sqrt{A})-f(\sqrt{E^{\min}}),
\end{align*}
\begin{align*}
\mathcal{D}(\tilde{\Sigma}^{\mathrm{t-m}}_\infty )=f(\sqrt{B})-f(\nu_-)-f(\nu_+)+f(\sqrt{E^{\min}}),
\end{align*}
with
\begin{align*}
&E^{\min}=\begin{cases}
\frac{2 C^2+(B-1)(D-A)+2|D|\sqrt{C^2+(B-1)(D-A)}}{(B-1)^2} , & \\ \qquad \textrm{for} \quad (D-AB)^2 \leq (1+B)C^2(A+D) \\
\frac{AB-C^2+D-\sqrt{C^4+(-AB+D)^2-2C^2(AB+D)}}{2B}, & \\ \qquad \textrm{otherwise}
\end{cases}
\end{align*}
and 
\begin{align*}
    f(x)=\bigg( \frac{x+1}{2} \bigg) \log \bigg[ \frac{x+1}{2} \bigg]-\bigg( \frac{x-1}{2} \bigg) \log \bigg[ \frac{x-1}{2} \bigg].
\end{align*}
We used that, according to Williamson's theorem, $\nu_+$, $\nu_-$ are the pairwise occuring \textit{symplectic eigenvalues} of the covariance matrix, obtained as the diagonal elements of the symplectic diagonalized matrix $2\tilde{\Sigma}^{t-m}_{\infty}$.\\
Based on Eq.~\eqref{CMTwoMode} one can also study entanglement in the system. In terms of the smallest symplectic eigenvalue $\nu_-$, a physically permissible (bonafide) covariance matrix  has to fulfill $\nu_-\geq1$. The violation-degree of this condition under partial transposition of the underlying density matrix can be quantified by the \textit{logarithmic negativity}
\begin{align*}
\mathcal{E}_{\mathcal{N}}=\max (0,-\log ( \tilde{\nu}_- )).
\end{align*}
Here $\tilde{\nu}_-$ is the smallest symplectic eigenvalue of the partially transposed covariance matrix, obtained from $2\tilde{\Sigma}^{t-m}_{\infty}$ by exchanging $F^p\to -F^p$. Non-zero values of $\mathcal{E}_{\mathcal{N}}$ are necessary and sufficient for entanglement (non-separability) between the spin ensemble and the bosonic mode as a result of the Peres-Horodecki criterion. \\

\subsection{Stationary quantum discord and classical correlation for measurements on the spin system}
As already mentioned in the main text, the quantum discord and the classical correlation allow for analog definitions with respect to measurements on the spins. Within the discussion of Gaussian states, in terms of covariance matrices, these definitions can be achieved by accordingly interchanging $\det (2\Gamma_s) \leftrightarrow \det (2\Gamma_b)$ in the formulae given above. 

\begin{figure}[h]
\centering
    \includegraphics[width=0.55\linewidth]{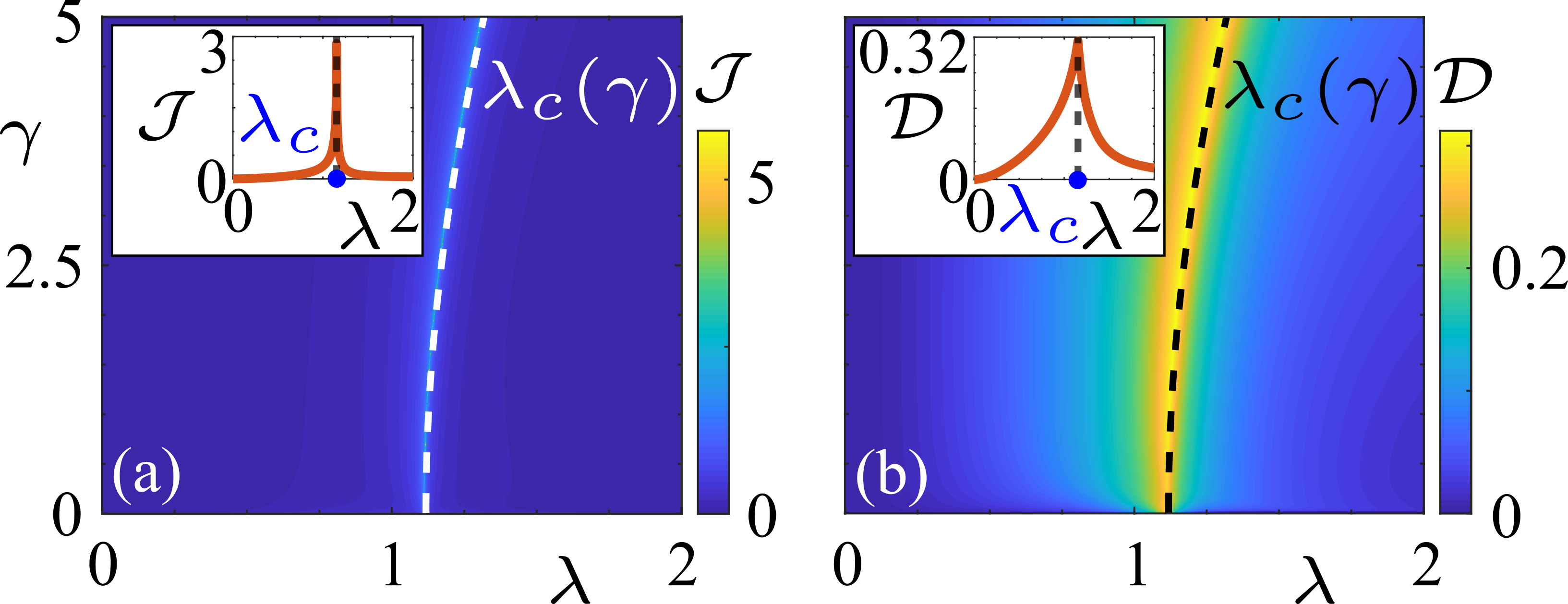}
    \caption{\textbf{Quantum discord and classical correlation for spin-measurements.} (a) Classical correlation $\mathcal{J}$ and (b) quantum discord $\mathcal{D}$ as functions of $\gamma$ and $\lambda$. Here $\mathcal{J}$ was maximized over all POVMs on the spin system. The insets visualize the $\lambda$-dependence of the respective quantities for $\gamma=2$. The plots were produced assuming that $\omega_m=1$ and $\omega_z=4$. All parameters are in units of $\kappa$.}
    \label{QDCC_otherway}
\end{figure}

The parameter dependence of $\mathcal{J}$ and $\mathcal{D}$ in this case is illustrated in Fig.~\ref{QDCC_otherway}. We can see from this figure that the discord has now maxima of approximately the same height, still distributed along the critical line $\lambda_c(\gamma)$. In contrast, the classical correlation shows essentially the same behavior as for the case where the measurements were performed on the boson system. It still diverges at the critical line (in the plots the asymptotic value is bounded by the chosen parameter resolution).
\end{document}